\documentclass[prl,twocolumn,showpacs,preprintnumbers,amsmath,amssymb, superscriptaddress]{revtex4-1}
\usepackage{graphicx}
\usepackage{dcolumn}
\usepackage{bm}
\usepackage{fancyhdr}
\usepackage{float}
\usepackage{ifthen}
\usepackage{eurosym}
\usepackage{wrapfig}
\usepackage{xfrac}
\usepackage{color}

\usepackage[normalem]{ ulem }
\usepackage{soul}

\begin{document}
\title{Kondo-induced giant isotropic negative thermal expansion}

\ \author{D.~G.~Mazzone}
\ \email{daniel.mazzone@psi.ch}
\ \affiliation{National Synchrotron Light Source II, Brookhaven National Laboratory, Upton, New York 11973, USA}
\ \affiliation{Condensed Matter Physics and Materials Science Department, Brookhaven National Laboratory, Upton,
New York 11973, USA}

\ \author{M.~Dzero}
\ \affiliation{Department of Physics, Kent State University, Kent, OH, 44242, USA}

\ \author{AM.~M.~Abeykoon}
\ \affiliation{National Synchrotron Light Source II, Brookhaven National Laboratory, Upton, New York 11973, USA}

\ \author{H.~Yamaoka}
\ \affiliation{RIKEN SPring-8 Center, Sayo, Hyogo 679-5148, Japan}

\ \author{H. Ishii}
\ \affiliation{National Synchrotron Radiation Research Center, Hsinchu 30076, Taiwan}

\ \author{N.~Hiraoka}
\ \affiliation{National Synchrotron Radiation Research Center, Hsinchu 30076, Taiwan}

\ \author{J.-P.~Rueff}
\ \affiliation{Synchrotron SOLEIL, L'Orme des Merisiers, BP 48 Saint-Aubin, 91192 Gif sur Yvette, France}
\ \affiliation{Sorbonne Universit\'{e}, CNRS, Laboratoire de Chimie Physique-Mati\`{e}re et Rayonnement, 75005 Paris, France}

\ \author{J.~M.~Ablett}
\ \affiliation{Synchrotron SOLEIL, L'Orme des Merisiers, BP 48 Saint-Aubin, 91192 Gif sur Yvette, France}

\ \author{K.~Imura}
\ \affiliation{Department of Physics, Nagoya University, Nagoya 464-8602, Japan}

\ \author{H. S. Suzuki}
\ \affiliation{Research Center for Advanced Measurement and Characterization, National Institute for Materials Science (NIMS), Sengen, Tsukuba 305-0047, Japan}
\ \affiliation{The Institute for Solid State Physics, The University of Tokyo, Kashiwanoha, Kashiwa 277-8581, Japan}

\ \author{J.~N.~Hancock}
\ \affiliation{Department of Physics and Institute for Materials Science, University of Connecticut, Storrs, Connecticut 06269, USA}

\ \author{I.~Jarrige}
\ \email{jarrige@bnl.gov}
\ \affiliation{National Synchrotron Light Source II, Brookhaven National Laboratory, Upton, New York 11973, USA}

\date{\today}
             
\begin{abstract}
Negative thermal expansion is an unusual phenomenon appearing in only a handful of materials, but pursuit and mastery of the phenomenon holds great promise for applications across disciplines and industries. Here we report use of X-ray spectroscopy and diffraction to investigate the 4$f$-electronic properties in Y-doped SmS and employ the Kondo volume collapse model to interpret the results. Our measurements reveal an unparalleled decrease of the bulk Sm valence by over 20$\%$ at low temperatures in the mixed-valent golden phase, which we show is caused by a strong coupling between an emergent Kondo lattice state and a large isotropic volume change. The amplitude and temperature range of the negative thermal expansion appear strongly dependent on the Y concentration and the associated chemical disorder, providing control over the observed effect. This finding opens avenues for the design of Kondo lattice materials with tunable, giant and isotropic negative thermal expansion.
\end{abstract}

\maketitle
Paradigms in the physics of correlated materials are often born from consideration of either localized or itinerant states, placing intermediate cases at the boundaries of conventional behavior. For example, isostructural valence transitions typified by metallic cerium (Ce) and samarium sulfide (SmS) are characterized by a strong change in the count of localized \emph{f}-electrons and therefore straddle the divide between these conceptual limits. In these systems, the role of itinerant \emph{spd} electronic degrees of freedom has been a central point of divergence between the two leading theoretical approaches, as the itinerant electronic degrees of freedom are crucial for the Kondo-volume-collapse model (KVC) \cite{Allen1982} while they remain essentially irrelevant in the Mott localization transition picture \cite{Johansson1974}. 

SmS is a black-colored semiconductor which undergoes a remarkable first-order volume collapse into a gold-colored metallic state under hydrostatic pressure or chemical substitution of Sm with Y \cite{Jayaraman1970, Maple1971, Coey1976}. Golden Sm$_{1-x}$Y$_{x}$S ($x$ $>$ 17\%) has recently garnered renewed focus, in particular for a potential topological transition in the Fermi surface as the intermediate valence monotonically decreases upon cooling \cite{Sousanis2017, Li2014, Matsubayashi2007, Kang2015, Corder2017, Kang2019}. Adding further interest to this material are reports of an exceptional occurrence of strong, isotropic, thermally robust and tunable negative thermal expansion (NTE) \cite{Hailing1984, Schefzyk1986, Takenaka2019}. Although discoveries of materials with strong NTE over wide temperature ranges have been greatly accelerated during the past two decades, thereby boosting the potential for integration in state-of-the-art technologies \cite{Monroe2018,Goel2018}, detailed understanding of the underpinning mechanisms, and the extent to which they can be enumerated is still under debate.

In this Letter, we set out to investigate the interplay between the electronic and structural properties in golden Sm$_{1-x}$Y$_{x}$S and their implication in the occurrence of NTE. To this end, we used advanced synchrotron X-ray tools to precisely determine the $f$-orbital occupation number $n_{f}$ and the lattice constant as a function of Y-concentration and temperature. Aided by a theoretical analysis based on the Kondo volume collapse (KVC) and the two-fluid model \cite{Nakatsuji2004}, we show that this material is characterized at low temperatures by an unusually strong increase in $n_{f}$ which couples with the lattice through a KVC. The effect culminates in a giant NTE, contrary to elemental Ce where the KVC model was originally developed to describe the strong positive thermal expansion. The Y substitution produces first-order valence and lattice transitions locally but an effectively broadened transition on macroscopic scales, and ultimately weakens the NTE behavior as these macroscopic effects are exacerbated for higher Y concentrations. The discovery of an NTE driven by hybridization between \emph{f} and $spd$-orbitals suggests that materials with hybridization between predominantly localized and itinerant electronic degrees of freedom provide promising opportunities for the realization of a new class of single-phase, isotropic and iso-structural giant NTE systems, tunable through application of pressure and/or alloying.


Single crystalline Sm$_{1-x}$Y$_{x}$S with $x$ = 0, 0.03, 0.14, 0.23 and 0.33 were synthesized by a Bridgman method \cite{Imura2009}, for which the actual Y concentration was determined using an inductively-coupled-plasma atomic-emission technique \cite{Imura2015}. High-resolution X-ray absorption spectroscopy was performed at the Sm $L_3$-edge in the partial fluorescence yield mode (PFY-XAS) by monitoring the intensity emitted over a 1 eV window around the maximum of the $L\alpha_{1}$ emission line while scanning the incident energy through the Sm $L_{3}$ edge. The experiment was carried out at the Taiwan beamline BL12XU in SPring-8 using the same experimental setup as in Ref. \cite{Jarrige2013}, and using a similar setup at the GALAXIES beamline at the SOLEIL synchrotron \cite{galax1, galax2}. X-ray podwer diffraction measurements were performed at the 28-ID-1 (PDF) beamline at NSLS-II. The unit cell volumes were refined in space group $Fm\bar{3}m$ as a function of temperature by means of a Rietveld and Le Bail analysis using the GSAS-II software package \cite{Gsas2013}.

\begin{figure}[tbh]
\includegraphics[width=\linewidth]{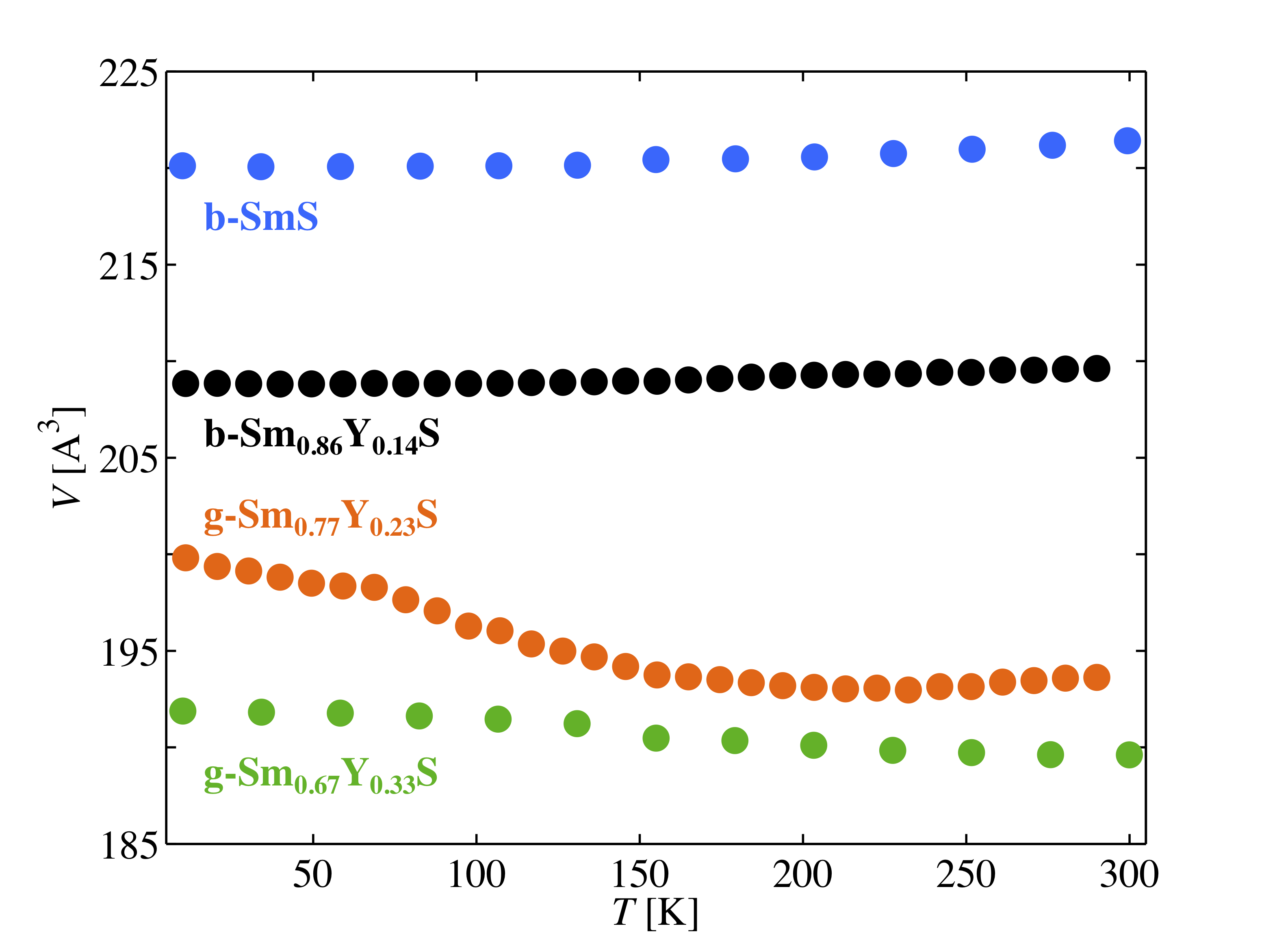}
\caption{(Color online) Temperature dependent unit cell volume of  $b-$SmS, $b-$Sm$_{0.86}$Y$_{0.14}$S,  $g-$Sm$_{0.77}$Y$_{0.23}$S and  $g-$Sm$_{0.67}$Y$_{0.33}$S.}
\label{fig1}
\end{figure}

Figure \ref{fig1} shows the temperature dependence of the unit cell volume of unsubstituted $b-$SmS, the black-colored 14$\%$ system $b-$Sm$_{0.86}$Y$_{0.14}$S, the golden 23$\%$ compound $g-$Sm$_{0.77}$Y$_{0.23}$S and $g-$Sm$_{0.67}$Y$_{0.33}$S. While the unit cell volumes of $b-$SmS and $b-$Sm$_{0.86}$Y$_{0.14}$S show conventional positive thermal expansion with a relative change of $\sim$0.6\% between 10 and 300 K, NTE is observed in the golden phase, with marked increases in the unit cell volume of 3.3\% and 1.5\% for $g-$Sm$_{0.77}$Y$_{0.23}$S and $g-$Sm$_{0.67}$Y$_{0.33}$S respectively, in agreement with a recent study \cite{Takenaka2019}.
\bigskip

\begin{figure}[tbh]
\includegraphics[width=\linewidth]{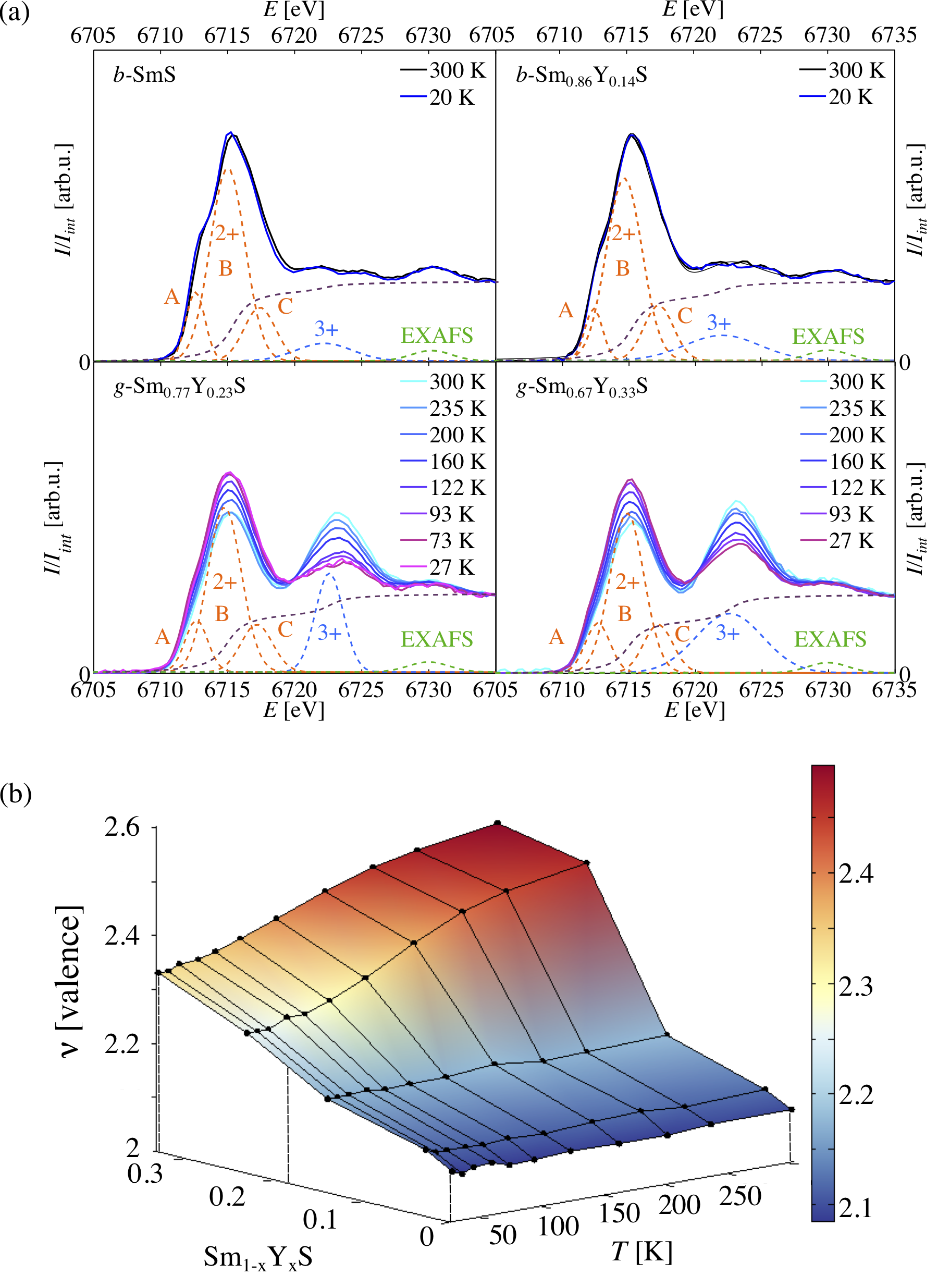}
\caption{(Color online) \textbf{a)} Temperature dependence of the Sm-L$_3$ PFY-XAS spectra for $b-$SmS, $b-$Sm$_{0.86}$Y$_{0.14}$S, $g-$Sm$_{0.77}$Y$_{0.23}$S and $g-$Sm$_{0.67}$Y$_{0.33}$S, respectively.  The dashed lines display the best fit to the base temperature data. \textbf{b)} Temperature and doping dependence of the Sm valence.}
\label{fig2}
\end{figure}
\bigskip

The microscopic origin of NTE in golden SmS was studied via PFY-XAS, which enables us to estimate the temperature and Y-substitution dependence of the Sm valence with a higher accuracy than conventional X-ray absorption spectroscopy \cite{Jarrige2013}. The experimental data, shown in Figure \ref{fig2}a reveal two main peaks corresponding to 2$p$ $\rightarrow$ 5$d$ dipolar excitations for divalent and trivalent Sm states. Each spectrum was fitted to three Gaussian functions for the divalent peak, one for the trivalent peak, one for the extended X-ray absorption fine structure, and an arctangent contribution accounting for the transitions to the continuum. As discussed in detail in Ref. \cite{Jarrige2013}, the divalent component requires a fit of three Gaussian contributions in order to satisfactorily reproduce the preshoulder A, the main peak B and the high-energy shoulder C, arising from crystal-field split 5$d$ $t_{2g}$ and $e_g$ manifolds. The exact Sm valence is derived from the relative intensity between the divalent and trivalent peak. 

Nearly invariant in the black phase, the valence undergoes a sharp increase right at the black-to-gold transition between $x$ = 0.14 and 0.23, which is expected from the volume collapse. The key experimental result of this study is the observation of a gradual but pronounced decrease of the valence upon cooling in the golden phase, down to $v$ = 2.26 in $g-$Sm$_{0.77}$Y$_{0.23}$S, indicating that, remarkably, the system reverts to a more divalent state at low temperatures. The valence decrease is noticeably weaker in $g-$Sm$_{0.67}$Y$_{0.33}$S, for which the low-temperature value is $v$ = 2.32. The behavior of the valence in the golden phase seems to be directly linked to NTE, as the system with the largest valence change, $g-$Sm$_{0.77}$Y$_{0.23}$S, also shows the largest NTE effect.

The valence decrease by $\sim$0.2 in $g-$Sm$_{0.77}$Y$_{0.23}$S at low temperatures is stronger by about an order of magnitude than the typical valence change displayed by Kondo materials, which are most often Ce or Yb-based  \cite{Fanelli2014, Eugene2018, Cornelius2002, Lawrence2001, Kummer2018}. Particularly interesting is the comparison with an Sm-based Kondo system that has recently drawn a lot of interest, SmB$_6$, and which is also in the strongly intermediate-valence regime. SmB$_6$ features a nearly zero thermal expansion while the valence decreases only by $\sim$0.07 electrons, which is considered quite strong in comparison to many Kondo-screened materials \cite{Matsubayashi2007, Phelan2014, Riseborough200, Mizumaki2009, Penney1974, Wachter1985, Jiao2016, Mizumaki2009}. This comparison further hints at an additional effect in low-temperature golden Sm$_{1-x}$Y$_{x}$S, beyond the usual Kondo crossover, contributing to the unconventional strong electron localization disclosed by the PFY-XAS data. 

In fact, the strong coupling between the large changes in the charge and lattice degrees of freedom in Sm$_{1-x}$Y$_{x}$S is reminiscent of the KVC which underlies the $\alpha$-$\gamma$ phase transition in elemental Ce \cite{Allen1982}. The phase transition in Ce is characterized by a strong increase of the Kondo temperature concomitant with a volume collapse under pressure or upon cooling \cite{Rueff2006, Lipp2008, Decremps2011, Maxim2006, Lipp2012, Chiu2018}. This motivates us to attempt a description of the temperature-induced changes of the properties of golden Sm$_{1-x}$Y$_x$S using the KVC.

\begin{figure*}[tbh]
\includegraphics[width=\textwidth]{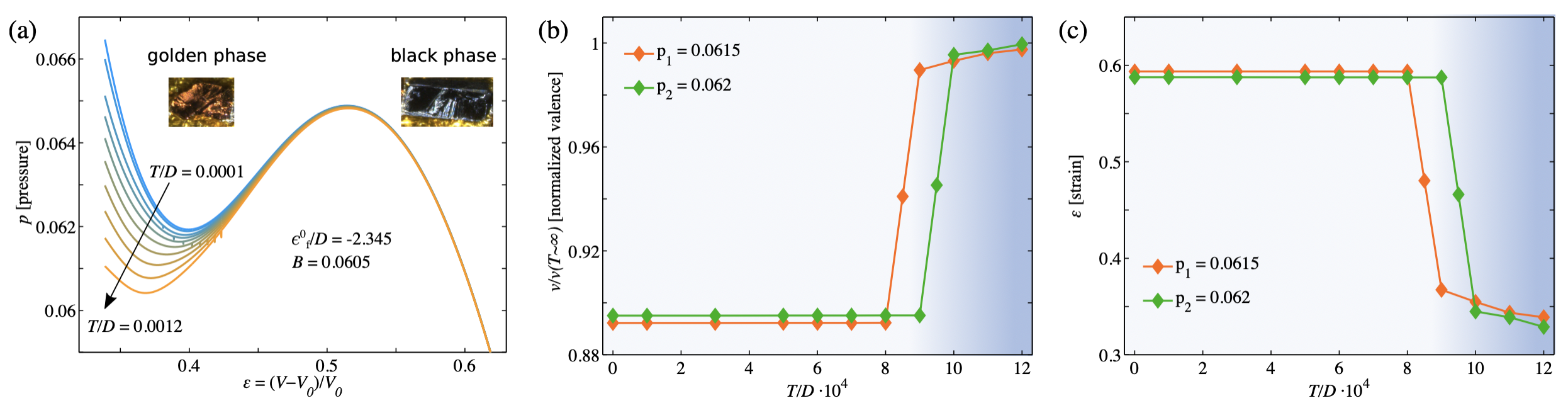}
\caption{(Color online) \textbf{a)} KVC equation of state for various reduced temperatures $T/D$. \textbf{b)} and \textbf{c)} Temperature dependent first-order valence, $\nu$, and volume transitions at pressures $p_{1}$ $<$ $p_{2}$.}
\label{fig3}
\end{figure*}
\bigskip

The physics of the KVC can be modeled by a Kondo lattice Hamiltonian, where the Anderson lattice, 
\begin{equation}
H_{AL} = \sum_{\vec{k},s}\epsilon_{\vec{k}}c_{\vec{k},s}^{\dagger}c_{\vec{k},s} + \sum_{i,s}\epsilon_{f}^0f_{i,s}^{\dagger}f_{i,s} + \Gamma\sum_{i,s}(f_{i,s}^\dagger c_{i,s} + h.c)\nonumber,
\end{equation}
is extended by $H_{lat}$ that describes the coupling between $f$-hybridization, $\Gamma$, and the local strain $\varepsilon_i$ \cite{Maxim2006}:
\begin{equation}
H_{lat} = -\gamma\Gamma\sum_{i,s}\varepsilon_i(f_{i,s}^\dagger c_{i,s} + h.c) +\frac{B}{2}\frac{V_0}{N_0}\sum_i\varepsilon_i^2.
	\label{Halt}
\end{equation}
$\gamma$ describes the linear volume dependence of the hybridization, $B$ is the bare bulk modulus, $V_0$ is a reference volume and $N_0$ denotes the number of unit cells in the system. $f_{i,s}^{(\dagger)}$ and $c_{i,s}^{(\dagger)}$ are the annihilation (creation) operators of an $f$- and conduction electron at position $i$ with spin $s$. Using a slave boson mean-field theory the following equation of state is derived for the strain-dependent pressure  $p$($\varepsilon$) \cite{Maxim2006}:

\begin{equation}
p(\varepsilon) = -B\varepsilon-2\gamma\frac{(\epsilon_f-\epsilon_f^0)(1-n_f)}{1-\gamma\varepsilon}.
	\label{eqs}
\end{equation}
Here, $\epsilon_f^0$ and $\epsilon_f$ are the bare and renormalized $f$-electron energy, respectively and $\varepsilon$ = ($V$-$V_0$)/$V_0$. The equation of state shows that the Kondo contribution to the pressure is negative if $\gamma$ is positive. We display $p(\varepsilon)$ in Fig. \ref{fig3}a for various reduced temperatures $T/D$, where $D$ is the conduction electron band width. The curves show the archetypal liquid-gas volume contraction under compression, which is strongly reminiscent of the $\alpha$-$\gamma$ KVC in Ce \cite{Allen1982, Lipp2008, Decremps2011}. Thus, the equation of state of Sm$_{1-x}$Y$_x$S properly describes the first-order volume collapse induced by chemical pressure, if we identify the theoretical predicted critical pressure with the experimentally determined critical Y concentration $x$ = 0.17.

In contrast to elemental Ce, it is the contracted phase in SmS that is Kondo-screened at low temperature. As displayed in Fig. \ref{fig3}a, the pressure-volume curves show that a decrease in the temperature results in an increase of the volume of the system (see black arrow). This volume expansion reaches a first-order transition for a critical temperature $T_C$. In fact, minimizing the slave boson mean-field free-energy with respect to $n_f$ confirms a first-order valence transition at $T_C$ (see Fig. \ref{fig3}b).
\begin{figure}[tbh]
\includegraphics[width=\linewidth]{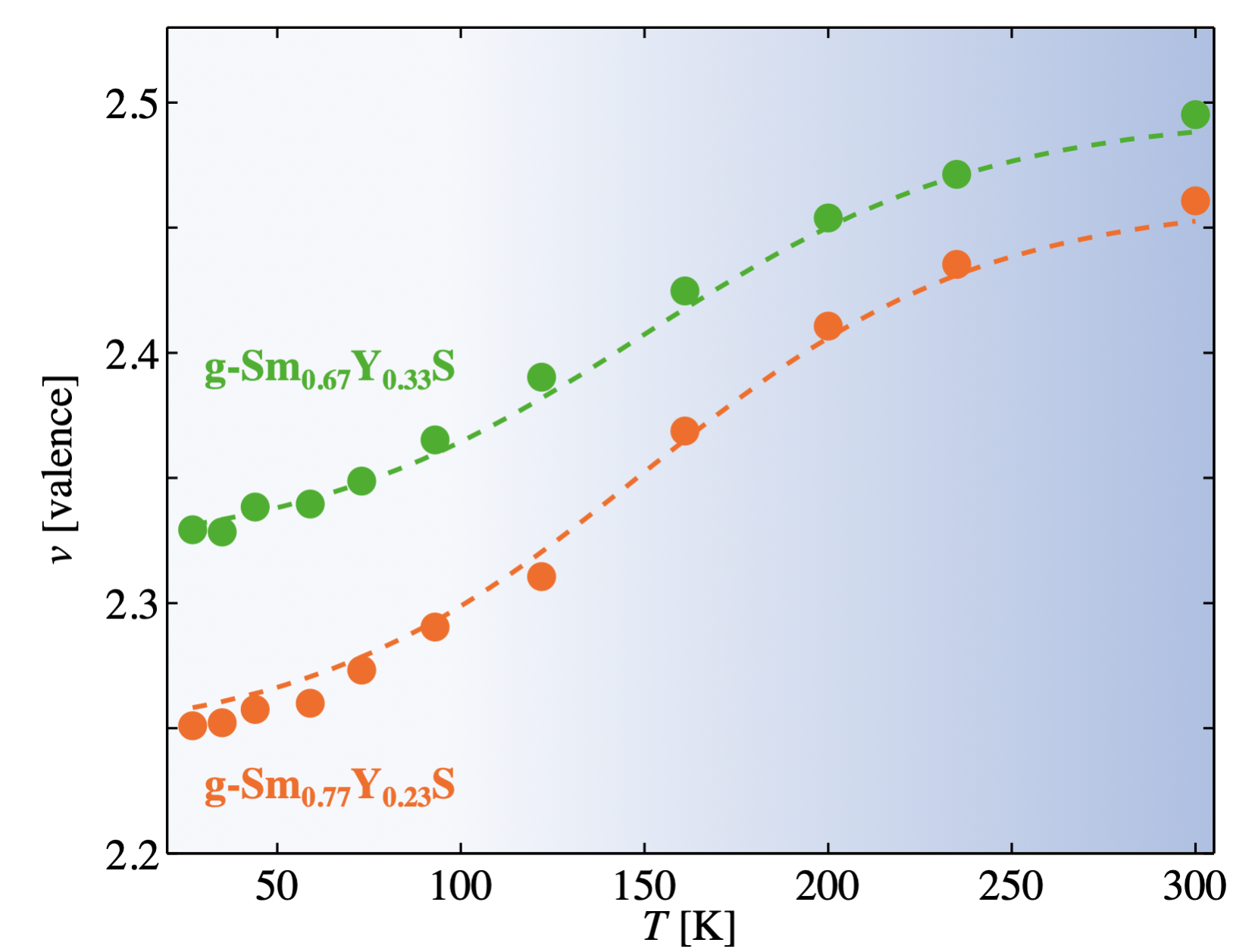}
\caption{(Color online) Fit of the experimental data (circles) with the two-fluid model (dashed line). The valence, $v$, is given as a function of  Y concentration, $x$, and temperature $T$: $v$($x$, $T$) = [$xv_{imp}$+(1-$x$)$v_{lat,HT}$]$f$($T_C$-$T$)+$v_{lat,LT}$$f$($T$-$T_C$) with $v_{imp}$ = 2.56, $v_{lat,LT}$ =  $v_{lat,HT}$ = 2.19 and $f$($T$) = [1 + $\tanh(\frac{T/T_C}{\sigma})$]/2 with $\sigma$ $\sim$ 0.615.} 
\label{fig4}
\end{figure}
\bigskip

The first-order nature of the temperature-induced phase transition derived from these thermodynamic calculations is, a priori, in stark contrast with the experimental observations of a continuous transition. A continuous transition can naturally arise from one of two mechanisms. In cases where the low-temperature critical line terminates at a critical pressure $p_c$, the temperature dependence becomes continuous at $p$ $>$ $p_c$ \cite{Maxim2006}. Alternatively, the first-order transition can be smeared out macroscopically due to local disorder, which is neglected in the Kondo lattice model. Results from previous experimental studies converge towards the latter scenario \cite{Menushenkov2006,Schefzyk1986,Takenaka2019,Tao1975}. Macroscopic susceptibility and dilatometry data, for instance, reveal an appreciable hysteresis in the temperature dependence of Sm$_{1-x}$Y$_{x}$S with $x$ close to 0.17, suggesting an $xT$-phase diagram in which the transition remains locally first order up to 0.27 $<$ $x_c$ $<$ 0.45 \cite{Takenaka2019,Tao1975}. Further, recent high-resolution X-ray diffraction results have shown that Sm$_{0.78}$Y$_{0.22}$S becomes phase separated at low temperatures, unravelling the coexistence of two isostructural domains featuring collapsed and extended lattice constants \cite{Takenaka2019}. This suggests that the NTE takes place via a droplet scenario, arising from variations in the local environment of the Sm atoms. Along these lines, we suggest to dilute the Kondo effect with increasing Y concentration. This can be described in a phenomenological two-fluid model, where two coexisting contributions of either residual local moments (fluid 1) or hybridized $f$-electrons (fluid 2) are assumed \cite{Yang2016, Nakatsuji2004}. A local order parameter, $f$($T$-$T_C$), allows to control how the volume fraction between the two fluids evolves as a function of temperature. A first-order transition that is broadened by a gradient of local disorder can be described by $f$($T$) = 1/2 + $\tanh[({T/T_C)/\sigma}$]/2. Thus, the temperature and doping dependence of the valence $v$ for our case reads:

\begin{align}
 v(x, T) &= [xv_{imp}+(1-x)v_{lat,HT}]f(T_C-T)\nonumber\\
 &+ v_{lat,LT}f(T-T_C)
	\label{tfm}
\end{align}
Here, $v_{imp}$ = 2.56 denotes the Sm valence in the completely diluted alloy (limit at large Y concentrations) and $v_{lat,HT}$ and $v_{lat,LT}$ are the Kondo lattice contributions at high ($T$ $\gg$ $T_C$) and low ($T$ $\ll$ $T_C$) temperatures. Since Sm$_{1-x}$Y$_{x}$S retains a temperature-independent mixed state for $x$ = 0, we set $v_{lat,LT}$ =  $v_{lat,HT}$ = 2.19. Best agreements to the experimental results at $x$ = 0.23 and 0.33 were found for a broadening factor $\sigma$ $\sim$ 0.615 and are shown in Fig. \ref{fig4} with $T_C$ = 145 and 150 K, respectively. The analysis implies that the experimental estimates of the valence reported in this work are averaged over coexisting black and golden domains, and thus differs from the intrinsically fluctuating Sm valence in a single phase.

Thus, the theoretical analysis of our experimental data strongly suggests that the electronic phase transitions in Sm$_{1-x}$Y$_{x}$S, both as a function of doping and temperature, arise from a KVC mechanism. Due to local disorder in the Y concentration the first-order character of the temperature-induced transition in golden SmS smears out macroscopically, leading to a continuous NTE over a $\sim$ 200 K range. The NTE is therefore enabled in this material through a combination of the Kondo screening of the local Sm $4f$ spins provoking an increase of the 4$f$-occupancy and the strong coupling between the electronic and lattice degrees of freedom. 

This scenario is in contrast with another theoretical model which assigns the phase transitions in SmS to an excitonic instability \cite{Kikoin1983, Mizuno2008, Imura2011}. The excitonic model predicts that compression leads to the softening of a bound electron-hole state, which ultimately condensates into the ground state at the black-to-gold transition. It is further claimed that in the golden phase the same exciton gaps out at low temperatures, triggering a pseudo-gap structure with semimetallic properties. However, we note that disorder induced into the system by the Y substitution is expected to impede the formation of a coherent excitonic wave-function, as a randomly distributed impurity potential locally breaks bound electron-hole pairs. Thus, a scenario based on exciton condensation is unlikely to explain the behavior in Y-doped SmS. It is also noted that recent dynamical mean-field theory study used a conventional broad Kondo crossover to describe the low-temperature behavior of golden SmS \cite{Kang2015}, but neglected the coupling to the crystal lattice. In our model it is this electron-lattice coupling that gives rise to a unified mechanism underpinning the pressure and temperature-induced volume collapses, and the associated giant NTE in the golden phase.

Finally, we discuss the possibility to find giant NTEs in other Kondo materials which are expected to show a valence decrease at low temperatures. Suitable candidates are limited to Sm, Tm, and Yb-based materials for lanthanides. While our results indicate that the proximity and, eventually, the coupling to a structural instability is a key ingredient, the example given by the large first-order valence decrease in YbInCu$_{4}$ at low temperatures can offer insight into the coupling of the Kondo effect to an electronic instability \cite{Jarrige2015}. In this system, despite a decrease of the valence by an amount nearly as large as $g-$Sm$_{0.77}$Y$_{0.23}$S ($\Delta n_{f}$ = 0.16), the volume change is 6.5 times smaller, at 0.5\%. This stresses the importance of the coupling between the Kondo effect and the lattice to generate a large NTE. Further, the weaker magnitude of the transition in SmB$_{6}$ shows that the relative number of rare-earth ions per unit cell needs to be substantial for the expansion of the ionic radius to have a sizeable effect on the lattice and provoke a volume collapse. This suggests that a combination of strong electron-lattice coupling and a large number of rare-earth ions per unit cell are both necessary ingredients to Kondo-induced NTE.

In summary, we have studied the structural and electronic properties of Sm$_{1-x}$Y$_{x}$S to identify the mechanism behind its giant isotropic NTE at $x$ $>$ 0.17. We find that it is directly linked to a massive decrease in the Sm valence, providing evidence for a strong coupling between the electronic and lattice degrees of freedom. A treatment of the experimental results in the KVC and two-fluid models provides a natural explanation for the instabilities induced by chemical pressure and temperature. In this scenario the temperature-induced volume collapse is broadened by chemical disorder, leading to a continuous NTE over 200 K. This study shows the potential of Kondo lattice materials for practical applications requiring tunable, isotropic NTE.

We acknowledge Priscila Rosa for fruitful discussions. D.G.M. acknowledges funding from the Swiss National Science Foundation, Fellowship No. P2EZP2\_175092. This research used the 28-ID -1 (PDF) beamline of the National Synchrotron Light Source II, a U.S. Department of Energy (DOE) Office of Science User Facility operated for the DOE Office of Science by Brookhaven National Laboratory under Contract No. DE-SC0012704. M. D. and J. N. H. acknowledge support from the U.S. Department of Energy, Basic Energy Sciences, Grant No. DE-SC0016481. M. D. acknowledges additional support from the National Science Foundation Grant No. NSF-DMR-1506547. J. N. H. acknowledges additional support from the National Science Foundation Grant No. NSF-DMR-1905862. The measurements of the PFY-XAS spectra were performed at the SPring-8 Taiwan beamline BL12XU under SPring-8 Proposal No. 2014B4270, corresponding NSRRC Proposal No. 2015-1-096, and at SOLEIL under Proposal No. 20171505.

\bibliography{main.bib}

\end{document}